\documentclass[reprint, superscriptaddress, aps, prx]{revtex4-2}

\usepackage{amsmath}
\usepackage{amssymb}
\usepackage{indentfirst}
\usepackage{commath}
\usepackage{graphicx}
\usepackage[caption=false,position=bottom,labelfont={bf}]{subfig}
\usepackage{overpic}
\usepackage{dcolumn}
\usepackage{bm}
\usepackage{booktabs}
\usepackage{dsfont}
\usepackage{setspace}
\usepackage{siunitx}
\usepackage{multirow}
\usepackage{tikz}
\usetikzlibrary{arrows.meta, positioning}

\usepackage{xcolor}
\definecolor{myred}{rgb}{0.8,0.1,0.2}
\definecolor{myblue}{rgb}{0.1,0.2,0.6}
\definecolor{relaxationblue}{RGB}{82,124,197}
\definecolor{quantumblue}{RGB}{145,169,211}
\definecolor{variationalblue}{RGB}{216,225,240}

\usepackage[colorlinks,linktocpage,hypertexnames=false]{hyperref}
\hypersetup{
    colorlinks=true,
    linktoc=all,
    linkcolor={myred},
    citecolor={myblue},
    urlcolor={myblue},
}

\usepackage{xparse}
\NewDocumentCommand{\bra}{o m}{
  \IfNoValueTF{#1}
    {\left\langle #2 \right|}
    {\langle #2 |}
}
\NewDocumentCommand{\ket}{o m}{
  \IfNoValueTF{#1}
    {\left| #2 \right\rangle}
    {| #2 \rangle}
}
\ExplSyntaxOn
\NewDocumentCommand{\expt}{O{true} m G{} G{}}{
  \str_if_eq:nnTF {#1} {true}
    {
      \tl_if_empty:nTF {#3}
        {\left\langle #2 \right\rangle}
        {
          \tl_if_empty:nTF {#4}
            {\left\langle #2 \middle| #3 \right\rangle}
            {\left\langle #2 \middle| #3 \middle| #4 \right\rangle}
        }
    }
    {
      \tl_if_empty:nTF {#3}
        {\langle #2 \rangle}
        {
          \tl_if_empty:nTF {#4}
            {\langle #2 | #3 \rangle}
            {\langle #2 | #3 | #4 \rangle}
        }
    }
}
\ExplSyntaxOff

\DeclareMathOperator{\tr}{Tr}


\begin{document}
%--------------------------------------------------------------------------

%--------------------------------------------------------------------------
\title{
Nullspace-guided Adaptive Bootstrap of Quantum Many-body Systems
}
%--------------------------------------------------------------------------

\author{Xu-Cheng Wang}
\affiliation{State Key Laboratory of Surface Physics, Fudan University, Shanghai 200433, China}
\affiliation{Center for Field Theory and Particle Physics, Department of Physics, Fudan University, Shanghai 200433, China}

\author{Yang Qi}
\email{qiyang@fudan.edu.cn}
\affiliation{State Key Laboratory of Surface Physics, Fudan University, Shanghai 200433, China}
\affiliation{Center for Field Theory and Particle Physics, Department of Physics, Fudan University, Shanghai 200433, China}
\affiliation{Collaborative Innovation Center of Advanced Microstructures, Nanjing 210093, China}
\affiliation{Hefei National Laboratory, Hefei 230088, China}

\date{\today}

%--------------------------------------------------------------------------
\begin{abstract}
%--------------------------------------------------------------------------
We introduce a nullspace-guided adaptive (NGA) bootstrap method
that improves the energy lower bounds of quantum many-body ground states by refining the bootstrap basis in a dynamic and incremental way.
At each iteration, the optimized moment matrix reveals a nullspace of saturated positivity directions,
which is intuitively interpreted as annihilators of the approximate ground-state subspace.
The NGA bootstrap then prunes operators with small nullspace leverage
and grows the basis along descendants of these null directions.
By applying the NGA bootstrap to the transverse-field Ising chain, we obtain nearly exact energy lower bounds
because the algorithm automatically discovers the eigenoperator structure in terms of Jordan--Wigner fermions from a minimal local bootstrap basis.
For the Hubbard chain, it improves upon state-of-the-art energy lower bounds by up to two orders of magnitude,
reaching errors ranging from $10^{-3}$ down to $10^{-5}$ in the strongly correlated regimes.
We further show that the NGA framework can be used to improve the certified two-sided bounds on general observables.
In addition, the bootstrap error decreases approximately as a power law with increasing computational resources.
These results suggest that our method provides a practical and scalable route toward accurate bootstrap of general quantum many-body systems.
%--------------------------------------------------------------------------
\end{abstract}
%--------------------------------------------------------------------------

\maketitle

\section{Introduction}
%--------------------------------------------------------------------------
Determining ground-state properties, especially the ground-state energy,
of interacting quantum many-body systems is a central challenge in quantum physics.
The difficulty stems from the exponential growth of the Hilbert space,
which makes exact diagonalization~\cite{weiße2008exact} limited to small systems and motivates a wide range of approximate methods.
Among them, quantum Monte Carlo~\cite{sandvik1999stochastic,blankenbecler1981monte} can be numerically exact when the sign problem is absent,
while variational approaches such as variational Monte Carlo~\cite{becca2017variational} and density-matrix renormalization group~\cite{schollwöck2011density} (DMRG)
provide a strict upper bound on the ground-state energy.

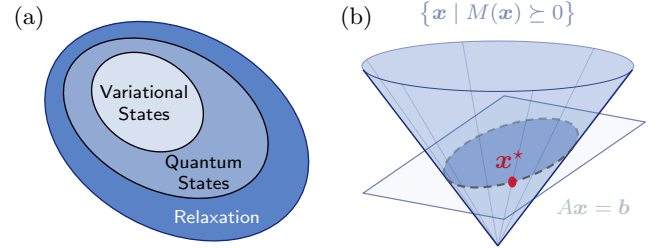
\begin{figure}[htbp]
    \centering
    \raisebox{-0.45cm}[0pt][0pt]{%
        \makebox[\columnwidth][l]{%
            \makebox[0.5\columnwidth][l]{\small (a)}%
            \makebox[0.5\columnwidth][l]{\small (b)}%
        }%
    }\par
    \begin{minipage}[c]{0.5\columnwidth}
        \centering
        \begin{tikzpicture}[
            scale=0.85,
            xscale=1.2,
            yscale=1.6,
            every node/.style={font=\sffamily\fontsize{7pt}{8pt}\selectfont},
            line join=round
        ]
            \begin{scope}[rotate=-15]
                \path[fill=relaxationblue, draw=myblue!80!black, line width=0.5pt, fill opacity=0.95]
                    (0,0) ellipse [x radius=1.75cm, y radius=1cm];
                \path[fill=quantumblue, draw=black, line width=0.5pt]
                    (-0.17cm,0.085cm) ellipse [x radius=1.36cm, y radius=0.72cm];
                \path[fill=variationalblue, draw=black, line width=0.5pt]
                    (-0.44cm,0.175cm) ellipse [x radius=0.74cm, y radius=0.46cm];
                \node[align=center] at (-0.48cm,0.16cm)
                    {Variational\\States};
                \node[align=center] at (0.44cm,-0.29cm)
                    {Quantum\\States};
                \node[text=white] at (0.72cm,-0.67cm)
                    {Relaxation};
            \end{scope}
        \end{tikzpicture}
    \end{minipage}%
    \hfill
    \begin{minipage}[c]{0.5\columnwidth}
        \centering
        % ========================== Adjustable parameters ==========================
        % Display and camera
        \def\xScale{1.2}
        \def\yScale{1.6}
        \def\camAz{45} % counterclockwise azimuth, in degrees
        \def\camEl{8}  % degrees above the u-v plane; larger -> more top-down
        % Affine plane: t = t0 + a u + b v
        \def\planeT{0.7}
        \def\planeU{0.27}
        \def\planeV{0.0}
        \def\patchU{1.1}       % displayed plane half-width in u
        \def\patchV{1.1}       % displayed plane half-width in v
        \def\patchCenterU{0.1} % displayed plane center in u
        \def\patchCenterV{0}   % displayed plane center in v
        % Cone display cap
        \def\capT{1.5}
        % Rendering
        \def\plotN{20}
        \def\capN{20}
        \def\planeOpacity{0.2}
        \def\feasibleOpacity{0.8}
        \def\coneOpacity{0.32}
        \def\capOpacity{0}
        \def\meshOpacity{0.2}
        \def\backOpacity{0.45}
        \def\frontOpacity{0.85}
        \def\meshAngles{60,120,180,240,300,360}
        % Labels and selected rank-one boundary point
        \def\starAngle{-120}
        \def\starDX{-0.35cm}
        \def\starDY{0.3cm}
        \def\capDX{0cm}
        \def\capDY{0.4cm}
        \def\planeDX{1.5cm}
        \def\planeDY{0cm}

        \begin{tikzpicture}[
            xscale=\xScale,
            yscale=\yScale,
            every node/.style={font=\sffamily\scriptsize},
            line join=round
        ]
            % M(t,u,v) = [[t+u,v],[v,t-u]], with M >= 0 iff t >= sqrt(u^2+v^2).
            % Orthographic camera projection:
            % (u,v,t) -> (X,Y) = (cos(a)u-sin(a)v, sin(e)[sin(a)u+cos(a)v]+cos(e)t).
            \pgfmathsetmacro{\xU}{cos(\camAz)}
            \pgfmathsetmacro{\xV}{-sin(\camAz)}
            \pgfmathsetmacro{\yU}{sin(\camEl)*sin(\camAz)}
            \pgfmathsetmacro{\yV}{sin(\camEl)*cos(\camAz)}
            \pgfmathsetmacro{\yT}{cos(\camEl)}
            \pgfmathsetmacro{\detXY}{\xU*\yV-\xV*\yU}
            \pgfmathsetmacro{\xR}{sqrt(\xU*\xU+\xV*\xV)}
            \pgfmathsetmacro{\xPhi}{atan(\xV/\xU)}
            \pgfmathsetmacro{\thetaA}{\xPhi+asin(-\detXY/(\yT*\xR))}
            \pgfmathsetmacro{\thetaB}{\xPhi+180-asin(-\detXY/(\yT*\xR))}
            \pgfmathsetmacro{\thetaC}{\thetaA+360}
            \pgfmathsetmacro{\starR}{\planeT/(1-\planeU*cos(\starAngle)-\planeV*sin(\starAngle))}

            % Finite patch of the affine plane t=t0+a*u+b*v.
            \path[fill=variationalblue, fill opacity=\planeOpacity,
                draw=myblue!80!black, draw opacity=0.60, line width=0.45pt]
                ({(\patchCenterU-\patchU)*\xU+(\patchCenterV-\patchV)*\xV},
                 {\yT*\planeT+(\patchCenterU-\patchU)*(\yT*\planeU+\yU)+(\patchCenterV-\patchV)*(\yT*\planeV+\yV)}) --
                ({(\patchCenterU+\patchU)*\xU+(\patchCenterV-\patchV)*\xV},
                 {\yT*\planeT+(\patchCenterU+\patchU)*(\yT*\planeU+\yU)+(\patchCenterV-\patchV)*(\yT*\planeV+\yV)}) --
                ({(\patchCenterU+\patchU)*\xU+(\patchCenterV+\patchV)*\xV},
                 {\yT*\planeT+(\patchCenterU+\patchU)*(\yT*\planeU+\yU)+(\patchCenterV+\patchV)*(\yT*\planeV+\yV)}) --
                ({(\patchCenterU-\patchU)*\xU+(\patchCenterV+\patchV)*\xV},
                 {\yT*\planeT+(\patchCenterU-\patchU)*(\yT*\planeU+\yU)+(\patchCenterV+\patchV)*(\yT*\planeV+\yV)}) -- cycle;

            % Feasible region: the affine slice inside the PSD cone.
            \path[fill=quantumblue, fill opacity=\feasibleOpacity]
                (0,{\planeT*\yT})
                plot[smooth, domain=0:360, samples=\capN, variable=\plotAngle]
                    ({\planeT*(\xU*cos(\plotAngle)+\xV*sin(\plotAngle))/(1-\planeU*cos(\plotAngle)-\planeV*sin(\plotAngle))},
                     {\planeT*(\yT+\yU*cos(\plotAngle)+\yV*sin(\plotAngle))/(1-\planeU*cos(\plotAngle)-\planeV*sin(\plotAngle))})
                -- cycle;

            % Cone surface through the t=const display cap.
            \path[fill=relaxationblue, fill opacity=\coneOpacity]
                (0,0) --
                plot[smooth, domain=\thetaA:\thetaB, samples=\plotN, variable=\plotAngle]
                    ({\capT*(\xU*cos(\plotAngle)+\xV*sin(\plotAngle))},
                     {\capT*(\yT+\yU*cos(\plotAngle)+\yV*sin(\plotAngle))})
                -- cycle;
            % Exact top interface: t=capT and u^2+v^2 <= capT^2.
            \path[fill=quantumblue, fill opacity=\capOpacity]
                plot[smooth, domain=0:360, samples=\capN, variable=\plotAngle]
                    ({\capT*(\xU*cos(\plotAngle)+\xV*sin(\plotAngle))},
                     {\capT*(\yT+\yU*cos(\plotAngle)+\yV*sin(\plotAngle))})
                -- cycle;

            % Generator rays and rim of the top interface.
            \foreach \plotAngle in \meshAngles {
                \draw[draw=myblue!80!black, draw opacity=\meshOpacity, line width=0.3pt]
                    (0,0) --
                    ({\capT*(\xU*cos(\plotAngle)+\xV*sin(\plotAngle))},
                     {\capT*(\yT+\yU*cos(\plotAngle)+\yV*sin(\plotAngle))});
            }
            \draw[draw=myblue!80!black, line width=0.55pt]
                (0,0) --
                ({\capT*(\xU*cos(\thetaA)+\xV*sin(\thetaA))},
                 {\capT*(\yT+\yU*cos(\thetaA)+\yV*sin(\thetaA))})
                (0,0) --
                ({\capT*(\xU*cos(\thetaB)+\xV*sin(\thetaB))},
                 {\capT*(\yT+\yU*cos(\thetaB)+\yV*sin(\thetaB))});
            \draw[draw=myblue!80!black, draw opacity=\backOpacity, line width=0.35pt]
                plot[smooth, domain=\thetaA:\thetaB, samples=\plotN, variable=\plotAngle]
                    ({\capT*(\xU*cos(\plotAngle)+\xV*sin(\plotAngle))},
                     {\capT*(\yT+\yU*cos(\plotAngle)+\yV*sin(\plotAngle))});
            \draw[draw=myblue!80!black, draw opacity=\frontOpacity, line width=0.35pt]
                plot[smooth, domain=\thetaB:\thetaC, samples=\plotN, variable=\plotAngle]
                    ({\capT*(\xU*cos(\plotAngle)+\xV*sin(\plotAngle))},
                     {\capT*(\yT+\yU*cos(\plotAngle)+\yV*sin(\plotAngle))});

            % Exact rank-deficient boundary det M=0 on the affine plane.
            \draw[densely dashed, draw=black!70, draw opacity=\backOpacity, line width=0.8pt]
                plot[smooth, domain=\thetaA:\thetaB, samples=\plotN, variable=\plotAngle]
                    ({\planeT*(\xU*cos(\plotAngle)+\xV*sin(\plotAngle))/(1-\planeU*cos(\plotAngle)-\planeV*sin(\plotAngle))},
                     {\planeT*(\yT+\yU*cos(\plotAngle)+\yV*sin(\plotAngle))/(1-\planeU*cos(\plotAngle)-\planeV*sin(\plotAngle))});
            \draw[densely dashed, draw=black!70, draw opacity=\frontOpacity, line width=0.8pt]
                plot[smooth, domain=\thetaB:\thetaC, samples=\plotN, variable=\plotAngle]
                    ({\planeT*(\xU*cos(\plotAngle)+\xV*sin(\plotAngle))/(1-\planeU*cos(\plotAngle)-\planeV*sin(\plotAngle))},
                     {\planeT*(\yT+\yU*cos(\plotAngle)+\yV*sin(\plotAngle))/(1-\planeU*cos(\plotAngle)-\planeV*sin(\plotAngle))});

            \coordinate (capCenter) at (0,{\capT*\yT});
            \coordinate (planeEdge) at
                ({(\patchCenterU-\patchU)*\xU+\patchCenterV*\xV},
                 {\yT*\planeT+(\patchCenterU-\patchU)*(\yT*\planeU+\yU)+\patchCenterV*(\yT*\planeV+\yV)});
            \node[anchor=south, xshift=\capDX, yshift=\capDY,
                text=quantumblue!85!black, font=\sffamily\footnotesize]
                at (capCenter) {$\left\{\bm{x}\mid M(\bm{x})\succeq0\right\}$};
            \node[anchor=west, xshift=\planeDX, yshift=\planeDY,
                text=variationalblue!85!black, font=\sffamily\footnotesize]
                at (planeEdge) {$A\bm{x}=\bm{b}$};
            \coordinate (xstar) at
                ({\starR*(\xU*cos(\starAngle)+\xV*sin(\starAngle))},
                 {\starR*(\yT+\yU*cos(\starAngle)+\yV*sin(\starAngle))});
            \fill[myred] (xstar) circle[radius=1.3pt];
            \node[anchor=west, xshift=\starDX, yshift=\starDY,
                text=myred!90!black, font=\sffamily\normalsize]
                at (xstar) {$\bm{x}^\star$};
        \end{tikzpicture}
    \end{minipage}
    \caption{%
        (a) The nesting of variational states, physical quantum states, and the relaxed feasible set.
        (b) Schematic illustration of the SDP feasible set in Eq.~\eqref{eq:sdp}.
        The optimum $\bm{x}^\star$ generically lies on the PSD cone boundary, where $M(\bm{x}^\star)$ has a nontrivial nullspace.
    }
    \label{fig:relaxation}
\end{figure}

The relaxation~\cite{coleman1963structure}, together with related bootstrap methods,
offers a complementary route to variational approaches
by minimizing the energy over a relaxed feasible set that contains all physical states.
As a result, it produces a rigorous lower bound on the ground-state energy, and together with variational approaches, can offer reliable estimations for quantum many-body systems.
The nesting of variational states, physical quantum states, and the relaxed feasible set is illustrated schematically in Fig.~\ref{fig:relaxation}(a).
In particular, the many-body bootstrap relaxes the full positivity condition of the density matrix.
One selects a finite bootstrap basis of operators and imposes positivity within the corresponding truncated operator space.
Additional symmetry constraints can also be imposed whenever they are expressible in terms of the retained operator moments.
This yields a semidefinite program~\cite{vandenberghe1996semidefinite,boyd2004convex} (SDP)
that optimizes over the independent expectation values of operator moments.
The size of the operator basis controls the size of the resulting SDP,
making the relaxation computable with finite resources.
This many-body bootstrap framework and similar relaxation ideas have found broad applications in
quantum chemistry~\cite{mazziotti2004realization,mazziotti2023quantum},
condensed-matter systems~\cite{
barthel2012solving,baumgratz2012lower,haim2020variational,wang2024certifying,kull2024lower,cho2025coarse,
hammond2006variational,verstichel2012variational,verstichel2013extensive,anderson2012second,
han2020quantum,scheer2026bootstrapping,gao2025bootstrapping,gao2026bootstrapping},
high-energy physics~\cite{han2020bootstrapping,berenstein2024numerical},
and quantum information~\cite{tavakoli2024semidefinite}.

However, the quality of the bootstrap bound depends crucially on the choice of the operator basis.
Conventional basis hierarchies use some simple truncation parameters,
such as the maximum degree or spatial range of operator strings,
which do not prioritize physically relevant operators in general
and lead to a rapid, combinatorial growth in the basis size.
Recent work has shown that the basis choice can be informed by the excitation spectrum.
In the symmetry-breaking phases of some spin models,
adding long-range string operators~\cite{chadha2026bootstrap,banerjee2026bootstrap},
or auxiliary link variables that locally encode the associated symmetry defects~\cite{scheer2025defect},
can effectively tighten bootstrap bounds.
Meanwhile, machine-learning approaches~\cite{requena2023certificates,flora2026moment} have also been explored
to guide the selection of bootstrap bases and constraints.
Nevertheless, a general, scalable, and physics-motivated strategy for adaptively improving the operator basis and associated bootstrap bounds remains lacking.

In this work, we introduce the nullspace-guided adaptive (NGA) bootstrap
that improves the bootstrap basis and the resulting energy bounds iteratively and incrementally.
The method uses the nullspace of the optimized moment matrix to identify operator directions
associated with the saturated positivity constraints,
which can be viewed heuristically as annihilators of the approximate ground-state subspace.
It then discards operators with less relevance to this nullspace
and grows the basis along missing descendants of the approximate annihilators.
For the transverse-field Ising chain, the NGA bootstrap automatically uncovers the intrinsic eigenoperator structure associated with the Jordan--Wigner fermions
and obtains a nearly exact energy bound.
Moreover, by applying NGA bootstrap to the Hubbard chain,
we improve upon state-of-the-art bootstrap bounds by typically one to two orders of magnitude,
reaching errors in the range of $10^{-5}$ to $10^{-3}$ in the strongly correlated regimes.
With increased computational resources, the bound accuracy exhibits favorable power-law scaling behavior.
In addition, we show that the NGA framework can systematically tighten the two-sided bounds on general observables.
Therefore, our method represents a concrete step toward accurate and scalable bootstrap of general many-body systems.
%--------------------------------------------------------------------------

\section{Bootstrap formulation}
%--------------------------------------------------------------------------
Given a finite-size many-body system with Hamiltonian $H$ and Hilbert space $\mathcal{H}$,
the many-body bootstrap aims to find a rigorous lower bound for the ground-state energy $E_0$.
It is based on the fundamental principle that a physical density matrix $\rho$ is positive semidefinite (PSD), $\rho\succeq 0$, and therefore
\begin{equation}
    \tr \left[\rho O^\dag O\right] \geq 0, \quad \forall\, O\in \mathcal{L}(\mathcal H),
\end{equation}
with $\mathcal{L}(\mathcal{H})$ the space of linear operators on $\mathcal{H}$.
Also, $\rho$ should be normalized, $\tr\rho=1$.
The bootstrap relaxation chooses a finite operator basis $\mathcal{B}=\{O_i:0\leq i<N\}$
and imposes positivity only on the subspace $\mathcal{V} = \mathrm{span}\,\mathcal{B} \subset \mathcal{L}(\mathcal{H})$,
\begin{equation}\label{eq:truncated_positivity}
    \expt{O^\dag O} \geq 0, \quad \forall\, O \in \mathcal{V}.
\end{equation}
$\expt[false]{\cdot}$ denotes a linear functional on $\mathcal{M}\to\mathbb{C}$,
where $\mathcal{M}=\mathcal{V}^\dag\mathcal{V}$ represents the moment operator space
spanned by $O_i^\dag O_j$ for $O_i,O_j\in\mathcal{V}$.
Since positivity is enforced only in the truncated space $\mathcal{V}$,
this functional need not extend to a positive functional on $\mathcal{L}(\mathcal{H})$
and therefore need not admit a density-matrix representation.
Equivalently, Eq.~\eqref{eq:truncated_positivity} requires the $N\times N$ moment matrix $M$ to be PSD, where
\begin{equation}
    M_{ij} = \expt{O^\dag_i O_j}.
\end{equation}
Hence $M\succeq 0$ realizes the relaxed \textit{positivity constraints} in the truncated operator space $\mathcal{V}$.
Let $\bm{x}$ denote the vector of independent expectation values (EVs) of operators in $\mathcal{M}$ that enter the moment matrix $M(\bm{x})$.
The optimization is performed over the vector $\bm{x}$, whose entries are SDP variables.
Provided that $H\in\mathcal{M}$, its expectation value can be written as $\expt{H}=\bm{c}^T\bm{x}$.
Then the minimization of $\expt[false]{H}$ subject to these constraints can be formulated as an SDP,
\begin{equation}\label{eq:sdp}
\begin{aligned}
    E_\text{SDP} = \min_{\bm{x}} \quad &\bm{c}^T\bm{x} \\
    \mathrm{s.t.} \quad &M(\bm{x}) \succeq 0,\, A\bm{x}=\bm{b}.
\end{aligned}
\end{equation}
The affine constraints $A\bm{x}=\bm{b}$ include the normalization condition $\expt{I}=1$ and \textit{symmetry constraints}.
If the density matrix respects a symmetry group $G$, represented by operators $S_g$ that may be unitary or antiunitary, then $S^{-1}_g\rho S_g=\rho$ for all $g\in G$.
We thus have symmetry constraints,
\begin{equation}
    \expt{O} =
    \begin{cases}
        \expt{S_g O S^{-1}_g}, & S_g \text{ unitary},\\[5pt]
        \expt{S_g O S^{-1}_g}^\ast, & S_g \text{ antiunitary},
    \end{cases}
\end{equation}
whenever both sides are representable in $\mathcal{M}$.
For a continuous unitary symmetry generated by a charge $Q$, this is infinitesimally realized as the Ward identity,
\begin{equation}
    \expt{\left[Q, O\right]} = 0.
\end{equation}
For example, thermal states and energy eigenstates are invariant under time translations,
yielding the stationarity condition $\expt[false]{[H,O]}=0$.
If $E_0$ is defined in a prescribed symmetry sector,
one may further impose sector constraints,
e.g. the fixed-particle-number sector $N=N_0$ can be enforced through
$\expt{N-N_0 I}=0$ and $\expt{(N-N_0 I)^2}=0$ in the truncated space $\mathcal{M}$.

Since the exact ground state must be feasible under the relaxed constraints in Eq.~\eqref{eq:sdp},
$E_\text{SDP}$ is a certified lower bound on the exact ground-state energy $E_0$,
\begin{equation}
    E_\text{SDP} \leq E_0.
\end{equation}
A nested enlargement of $\mathcal{B}$ reduces the feasible set of the relaxation,
therefore improving the energy lower bound toward $E_0$.

For lattice systems with translation symmetry, we choose a set of translation representatives
$
    \mathcal{B}_0=\{O_a\}_{a=1}^{N_a}
$.
Denote the translated operator $O_{a,r}=T^\dag(r) O_a T(r)$.
The full bootstrap basis $\mathcal{B}$ includes the complete translation orbits so that
$
    \mathcal{B} = \bigcup_{r} \mathcal{B}(r)
$
and
$
    \mathcal{B}(r) = \{O_{a,r}: 1\leq a \leq N_a\}
$.
This allows the moment matrix $M$ to be block diagonal in the momentum space,
thereby replacing the full PSD constraint by PSD constraints on smaller momentum blocks;
see Appendix~\ref{app:symmetries}.
%--------------------------------------------------------------------------

\section{Nullspace-guided adaptive bootstrap}
%--------------------------------------------------------------------------
The insight arises from the fact that
for a nontrivial SDP optimization,
the optimum is expected to lie on the boundary of the PSD cone,
as shown in Fig.~\ref{fig:relaxation}(b).
Otherwise, if the optimum were strictly inside the PSD cone,
Eq.~\eqref{eq:sdp} would locally degenerate to a linear optimization over an affine space,
which has no local minimum unless the affine constraints already fix the objective.
We exclude this degenerate case so that at the optimum $\bm{x}^\star$,
$M(\bm{x}^\star)$ supports a nontrivial nullspace.
The idea of NGA bootstrap is to iteratively examine these saturated positivity directions of $M(\bm{x}^\star)$, namely the moment nullspace.
Heuristically, this corresponds to refining the annihilators of the approximate ground-state subspace if the optimal moment functional is realizable by a density matrix.

First, we show that, within the operator space $\mathcal{V}$,
the moment nullspace identifies zero-norm operator directions induced by the optimized moment functional.
Let $\bm{v}_\alpha$ be a null vector of the optimized moment matrix,
\begin{equation}
    \bm{v}_\alpha^\dag\,M(\bm{x}^\star)\,\bm{v}_\alpha = 0,
\end{equation}
and define the corresponding operator in $\mathcal{V}$,
\begin{equation}
    P_\alpha = \sum_{a,r} \left(\bm{v}_\alpha\right)_{a,r} O_{a,r}.
\end{equation}
Combining them we have
\begin{equation}
    \expt{P_\alpha^\dag P_\alpha}_{\bm{x}^\star} = 0.
\end{equation}
Hence $P_\alpha$ represents a zero-norm operator direction in $\mathcal{V}$.
$\expt{\cdot}_{\bm{x}^\star}$ denotes the optimized moment functional $\mathcal{M}\to\mathbb{C}$.
One may formally represent it by an object $\rho^\star\in\mathcal{L}(\mathcal{H})$ so that
\begin{equation}
    \expt{\Omega}_{\bm{x}^\star} = \tr \left[\rho^\star \Omega\right], \quad \Omega\in\mathcal{M}.
\end{equation}
Note $\rho^\star$ is generally nonunique and need not be PSD.
In fact, a non-tight bound certifies that the optimal moments are pseudo-moments:
they cannot be extended to any density matrix in the target symmetry sector~\footnote{%
By definition, any density matrix lying strictly in the target symmetry sector has energy at least $E_0$.
If such a density matrix also reproduces the optimal moments, its energy would be $E_\text{SDP}$ and we have $E_\text{SDP}\leqslant E_0$ by relaxation.
Therefore, the optimal moment functional can be extended to a density matrix in the target symmetry sector only if the bound is exact.
For a non-tight bound, any normalized PSD realization of the optimal moment functional must have support outside the target symmetry sector.
}.
Nevertheless, we assume that the optimal moment functional admits a normalized PSD realization $\rho^\star$.
At finite truncation, it should be regarded as an effective density matrix
satisfying the imposed symmetry-sector constraints within $\mathcal{M}$
while not necessarily lying exactly in the symmetry sector with respect to the full operator algebra.
In the limit where $\mathcal{B}$ is enlarged to span the full operator algebra,
$\rho^\star$ can be chosen as the exact ground-state density matrix in the target symmetry sector.
With such a normalized PSD realization $\rho^\star$, the zero-norm condition implies that $P_\alpha$ annihilates the support of $\rho^\star$,
i.e. the approximate ground-state subspace,
\begin{equation}\label{eq:annihilator}
    P_\alpha \ket{\psi^\star} = 0, \quad
    \ket{\psi^\star}\in \mathrm{supp}(\rho^\star).
\end{equation}

This connection motivates the nullspace-guided update of the bootstrap basis.
The moment nullspace can be intuitively viewed as the annihilator space of the approximate ground-state subspace.
In other words, operator directions with negligible overlap with this nullspace contribute little to defining such annihilators,
i.e. they are irrelevant to the active positivity constraints.
Projecting out these directions is therefore expected to leave the energy bound nearly unchanged while reducing the size of the SDP.
In practice, exact projection breaks the sparsity of basis operators, which is unfavorable for both compiling and solving the SDP.
Alternatively, we calculate for each $O_a\in\mathcal{B}_0$ its leverage in the current moment nullspace,
\begin{equation}
    \ell_a = \frac{1}{d_0} \sum_{\alpha=1}^{d_0}\sum_{r=1}^{N_r}\, \abs{\left(\bm{v}_\alpha\right)_{a,r}}^2.
\end{equation}
Here, $d_0$ denotes the nullspace dimension and $N_r$ the number of lattice translations.
For orthonormal null vectors $\bm{v}_\alpha$, the leverage scores satisfy $\sum_a \ell_a = 1$.
All NGA moves are performed at the level of translation representatives $O_a$,
with each pruning or growing step applied simultaneously to its full translation orbits.
The leverage score measures how strongly the basis operator $O_a$ participates in the current nullspace.
Operators with small $\ell_a$ have little overlap with the approximate annihilator space and are therefore natural candidates for pruning.

From the dual perspective, the dual problem of Eq.~\eqref{eq:sdp} offers a sum-of-squares (SoS) proof of the energy lower bound:
under strong duality and complementary slackness~\cite{boyd2004convex},
the space spanned by nontrivial dual SoS operators is a subspace of the primal moment nullspace.
Within a perturbative framework, Hastings suggested building the SoS operators from
perturbatively dressed operators that approximately annihilate the ground state,
and the resulting certificates reproduced the associated perturbative energy corrections~\cite{hastings2024perturbation,hastings2024improving}.
In the same spirit, we use the moment nullspace to adaptively select a compact operator basis
that is expected to support a valid SoS certificate,
and the method is not restricted to the perturbative regime in general.

\begin{figure}[htbp]
    \centering
    \resizebox{0.8\columnwidth}{!}{
    \begin{tikzpicture}[
        node distance=0.45cm and 0.35cm,
        box/.style={
            draw,
            rounded corners,
            align=center,
            font=\scriptsize,
            inner sep=3pt,
            minimum width=2.0cm,
            minimum height=0.55cm
        },
        arrow/.style={-{Latex[length=1.6mm]}, semithick}
    ]
    \node[box] (init) {Initialize\\[1pt]$\mathcal{B}^{(0)}$};
    \node[box, below=of init] (solve) {Solve SDP};
    \node[box, below=of solve] (null) {Extract nullspace\\of $M(\bm{x}^\star)$};
    \node[box, below left=0.4cm and -0.5cm of null] (prune) {Prune by $\ell_a$};
    \node[box, below right=0.4cm and -0.5cm of null] (grow) {Grow by $s_\mu$};
    \node[box, below=1.35cm of null] (update) {Update\\[1pt]$\mathcal{B}^{(n+1)}$};

    \draw[arrow] (init) -- (solve);
    \draw[arrow] (solve) -- (null);
    \draw[arrow] (null) -- (prune);
    \draw[arrow] (null) -- (grow);
    \draw[arrow] (prune) -- (update);
    \draw[arrow] (grow) -- (update);
    \draw[arrow] (update.east) -- ++(1.85cm,0) |- (solve.east);
    \end{tikzpicture}
    }
    \caption{%
        Schematic workflow of the nullspace-guided adaptive bootstrap.
    }
    \label{fig:nga_flow}
\end{figure}
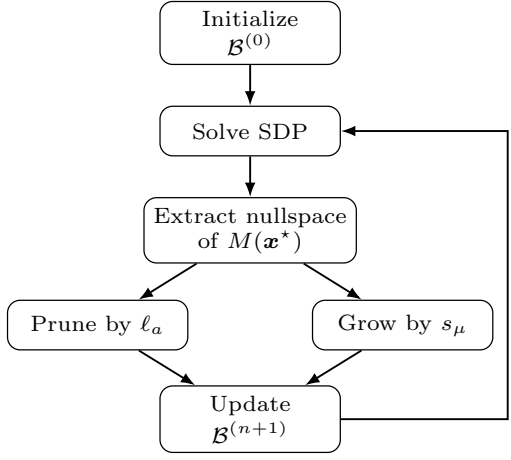

An intuitive way of growing the operator basis is to examine the relations in Eq.~\eqref{eq:annihilator} in an enlarged $\mathcal{V}$.
The guiding question is whether local variants of an approximate annihilator, such as its commutators with elementary operators, remain approximate annihilators.
In particular, we evaluate the Hamiltonian dynamics of $P_\alpha$, i.e. its commutator with the Hamiltonian,
\begin{equation}\label{eq:commutator}
    \left[H, P_\alpha\right]
\end{equation}
to generate such descendants of the null operators $P_\alpha$.
When the commutator contains components outside the current operator space $\mathcal{V}$,
those components provide natural candidates for the basis growth.
This is motivated by the following observations:
(i) For an exact ground-state density matrix and an exact ground-state annihilator,
the commutator Eq.~\eqref{eq:commutator} is again an exact ground-state annihilator.
Thus, if the current null operator $P_\alpha$ approximates an annihilator with respect to the effective $\rho^\star$,
its Hamiltonian descendants provide a natural probe to test and refine this annihilation relation.
(ii) The $H$-commutator serves as an efficient generator of descendants.
Because a local Hamiltonian contains terms with finite spatial range,
its commutator with a basis operator can both modify the operator content within the operator's existing spatial support and,
when a Hamiltonian term overlaps its support boundary, extend that support incrementally.
This balances the exploration of richer operator structures at fixed spatial support with the gradual growth of the support itself.
(iii) Hamiltonian descendants are directly related to the stationarity constraints, $\expt[false]{[H,O]}=0$.
Including operators generated by Eq.~\eqref{eq:commutator} helps close these constraints within the truncated moment space.

In particular, we consider
\begin{equation}
    \left(1-\Pi_{\mathcal{V}}\right) [H, P_\alpha] = \sum_{\mu,r} w_{\alpha,\mu,r} T^\dag(r) \tilde{O}_\mu T(r).
\end{equation}
$\Pi_{\mathcal{V}}$ denotes the projection onto the current operator space,
and $T^\dag(r) \tilde{O}_\mu T(r) \notin \mathcal{V}$ are potential operators for basis growth.
We assign each translation representative $\tilde{O}_\mu$ the score
\begin{equation}
    s_\mu = \frac{1}{d_0} \sum_{\alpha=1}^{d_0} \sum_{r=1}^{N_r}\, \abs{w_{\alpha,\mu,r}}^2.
\end{equation}
The candidates $\tilde{O}_\mu$ are then ranked by $s_\mu$,
and those with the highest scores are added to the bootstrap basis together with their full translation orbits.

In an NGA step, we prune operators that are irrelevant to the current moment nullspace
and grow the basis along missing descendants of the approximate annihilators.
Solving the SDP with the adaptive basis iteratively then provides increasingly tight lower bounds while maintaining a compact operator basis.
The typical workflow of NGA bootstrap is illustrated in Fig.~\ref{fig:nga_flow}.
We remark that the NGA framework here is general
while a dedicated growing strategy may offer further improvements for specific models.
%--------------------------------------------------------------------------

\section{Example: Ising chain}
%--------------------------------------------------------------------------
We first test the NGA bootstrap on the Ising chain with both transverse and longitudinal fields under periodic boundary conditions (PBC),
$
    H = -J \sum_i Z_i Z_{i+1} - h \sum_i X_i - h_z \sum_i Z_i
$.
We set $J=1$ as the energy unit; $h$ and $h_z$ denote the transverse and longitudinal field strengths.
The SDPs are solved with MOSEK optimizer~\cite{mosek} at default accuracy $\epsilon=10^{-8}$.
Accordingly, eigenvectors of the optimized moment matrix with eigenvalues below a threshold from $10^{-9}$ to $10^{-8}$
are identified as null vectors.
At each NGA step, we drop at most $5\%$ of the operators in the current basis $\mathcal{B}_0$,
while requiring a net increase of at least four representatives and no more than $5\%$ of its current size.
To discourage repeated reentry, the growth score $s_\mu$ of a previously removed operator is multiplied by $0.5$ for each prior removal.

\begin{figure}[htbp]
    \centering\hspace{-1cm}
    \includegraphics[width=.9\columnwidth]{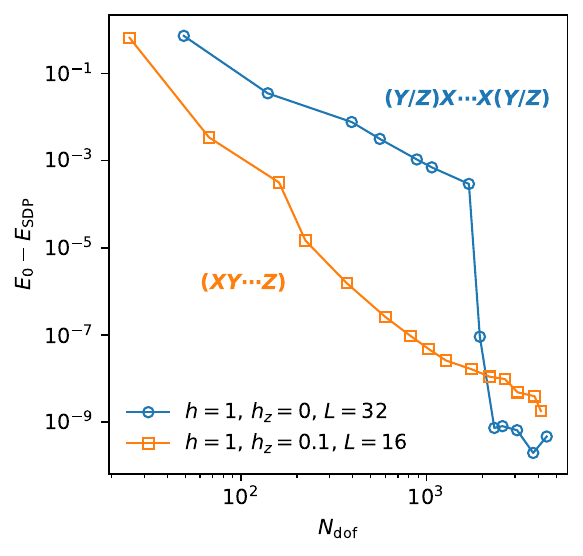}
    \caption{%
        NGA bootstrap of Ising chain.
        We denote by $N_{\mathrm{dof}}=\dim\bm{x}-\mathrm{rank}(A)$ the degrees of freedom of the SDP.
        For the transverse-field case at $L=32$, $h=1$, and $h_z=0$, the exact ground-state energy density $E_0$ is given by the JW solution.
        For the nonintegrable case at $L=16$, $h=1$, and $h_z=0.1$, $E_0$ is obtained from exact diagonalization.
        Both NGA runs start from the minimal basis $\mathcal{B}_0=\{I,X,Z\}$
        and terminate with $\abs{\mathcal{B}_0}=50$ for $h_z=0$ and $\abs{\mathcal{B}_0}=58$ for $h_z=0.1$.
    }
    \label{fig:ising}
\end{figure}

For $h_z=0$, the Ising chain is exactly solvable with Jordan--Wigner (JW) fermions and is critical at $h=1$~\cite{pfeuty1970one};
for $h_z\neq0$, the model is generally nonintegrable.
Fig.~\ref{fig:ising} shows the NGA bootstrap results for both cases.
Although the operator bases are not strictly nested for the NGA sequences that discard operators at each step,
the bounds still improve systematically.
At the integrable critical point $h=1$, $h_z=0$,
the bootstrap error decreases steadily with the NGA steps
and then drops sharply to the numerical precision.
This sudden improvement occurs when the adaptive basis discovers string operators with the JW structure,
schematically of the form $(Y/Z)X\cdots X(Y/Z)$, which are precisely the JW fermion bilinears.
This observation suggests that
once $\mathcal{V}$ sufficiently captures the annihilator structure of the exact ground-state subspace,
the SDP relaxation can attain the exact ground-state energy.
For the nonintegrable case $h=1$, $h_z=0.1$, the JW fermions are interacting and no analytic annihilator structure is known.
As a result, the adaptive growth prioritizes local operators, such as $XY\cdots Z$,
and the bootstrap error again decreases systematically with increasing SDP degrees of freedom.
We note that previous bootstrap studies~\cite{baumgratz2012lower,berenstein2024numerical} have shown that
tight bounds for the transverse-field Ising chain can be obtained by formulating the bootstrap natively in the JW fermion representation.
Here, starting from a minimal local basis,
our results demonstrate that the NGA bootstrap can automatically discover this intrinsic annihilator structure.
%--------------------------------------------------------------------------

\begin{figure*}[htbp]
    \centering
    \hspace{0cm}

    \begin{overpic}[percent,width=1\textwidth]{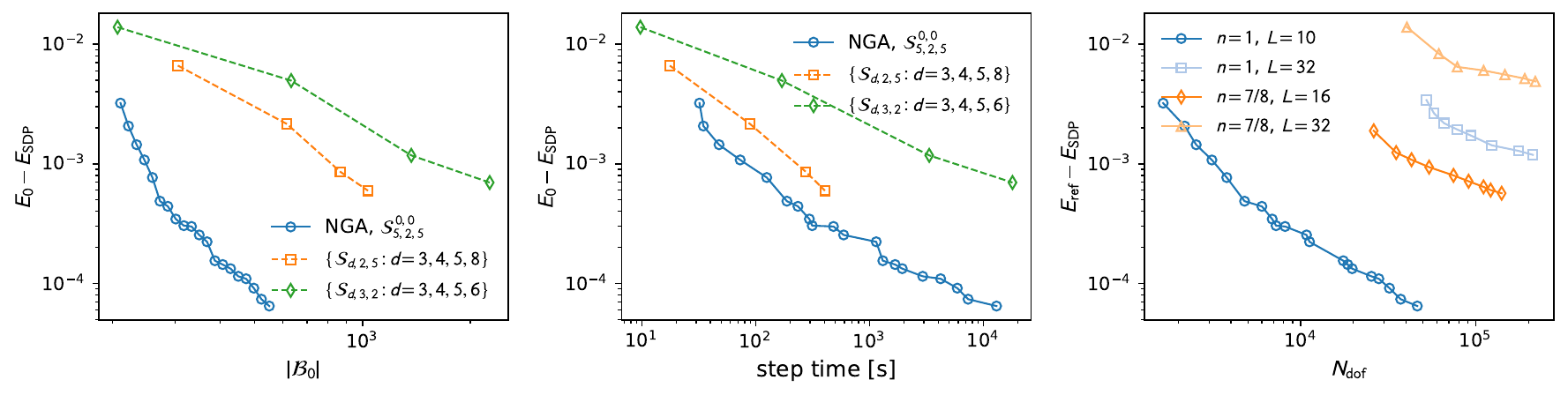}
        \put(-.5,22.5){\small\textbf{(a)}}
        \put(33,22.5){\small\textbf{(b)}}
        \put(66.4,22.5){\small\textbf{(c)}}
    \end{overpic}

    \vspace{.2cm}

    \makebox[0pt][l]{\hspace{-.04\textwidth}\raisebox{1.4cm}{\small\textbf{(d)}}}
    \begin{minipage}[htbp]{.9\textwidth}
    \begin{ruledtabular}
    \renewcommand{\arraystretch}{1.3}
    \begin{tabular}{cccccccc}
        $n$ & $L$ & $\mathcal{B}^{(0)}_0$ & $\abs{\mathcal{B}_0}$ & $E_\mathrm{SDP}$ & $E_\mathrm{ref}$ & $\Delta E$ & max step time\\
        \midrule
        1   & 10 & $\mathcal{S}^{0,0}_{5,2,5}$  & $211 \to 549$ & -1.58349690 & -1.583432263577  & $6.5\times 10^{-5}$ & $\sim 4\,\mathrm{h}$ \\
        1   & 32 & $\mathcal{S}^{0,0}_{6,2,8}$  & $456 \to 645$ & -1.57554402 & -1.5743563(1)    & $1.2\times 10^{-3}$ & $\sim 66\,\mathrm{h}$\\
        7/8 & 16 & $\mathcal{S}^{0,1}_{7,2,8}$  & $429 \to 638$ & -1.48039353 & -1.479826245187  & $5.7\times 10^{-4}$ & $\sim 30\,\mathrm{h}$\\
        7/8 & 32 & $\mathcal{S}^{0,1}_{3,2,16}$ & $381 \to 516$ & -1.47621495 & -1.471331(1)     & $4.9\times 10^{-3}$ & $\sim 75\,\mathrm{h}$\\
    \end{tabular}
    \end{ruledtabular}
    \end{minipage}
    \caption{%
        NGA bootstrap of Hubbard chain at $t=1$ and $U=4$.
        All energies are reported as energy densities with $\Delta E=E_\mathrm{ref}-E_\mathrm{SDP}$.
        The reference energies for $n=1$, $L=10$ and $n=7/8$, $L=16$ are obtained by exact diagonalization,
        while the other references are obtained by DMRG under PBC with $\chi_{\max}=4096$.
        For $n=1$ and $L=10$, we compare the NGA sequence and manually selected basis hierarchies,
        and show the energy bound gap versus (a) the basis size $\abs{\mathcal{B}_0}$ and (b) the computation time of each SDP step.
        (c)(d) NGA bootstrap results for $n=1$ and $n=7/8$ at different system sizes $L$.
        Detailed data are shown in (d), including the initial $\mathcal{B}_0$, the evolution of basis size $\abs{\mathcal{B}_0}$, the minimal energy bound gap, and the maximal per-step solving time.
        The $n=1$, $L=10$ run uses 4 CPU cores, while the other runs use 16 CPU cores,
        on an Intel Xeon Gold 6448H node with up to 1.5 TB memory allocated.
    }
    \label{fig:hubbard}
\end{figure*}

\section{Example: Hubbard chain}
%--------------------------------------------------------------------------
We next apply the NGA bootstrap to the Hubbard chain with PBC,
$
    H = -t \sum_{i,\sigma} (c^\dag_{i,\sigma}c_{i+1,\sigma}+c^\dag_{i+1,\sigma}c_{i,\sigma})
    + U \sum_i (n_{i\uparrow}-\frac{1}{2})(n_{i\downarrow}-\frac{1}{2})
$.
Here $t$ is the hopping amplitude, $U$ is the on-site interaction strength,
and we denote the electron filling by $n=N_0/L$.
We set $t=1$ and $U=4$, corresponding to a typical strongly correlated regime,
and consider both half filling $n=1$ and a doped case $n=7/8$.
The NGA parameters are chosen to be the same as those used for the Ising chain.
Technical details concerning the imposed constraints and symmetry reductions are provided in Appendix~\ref{app:hubbard}.

We formulate the Hubbard bootstrap in the Majorana representation.
Each operator in the bootstrap basis is an individual canonical Majorana monomial,
as defined in Eq.~\eqref{eq:majorana_monomial}.
For compact notation, we denote by
$
    \mathcal{S}_{d,s,r}
$
the set of translation-representative Majorana monomials with
degree up to $d$, support size up to $s$, and diameter up to $r$.
Here the degree is the number of Majorana fermions in the monomial,
the support size is the number of distinct lattice sites on which it acts,
and the diameter is the spatial extent of the shortest interval containing the occupied sites on the periodic chain.
This basis hierarchy can be further restricted to a fixed spin-resolved fermion-parity sector as
$
    \mathcal{S}^{p_\uparrow,p_\downarrow}_{d,s,r}
$.

The Hubbard model has been studied extensively in one and two dimensions through relaxations~\cite{
hammond2006variational,verstichel2012variational,verstichel2013extensive,anderson2012second,
han2020quantum,scheer2026bootstrapping}.
In particular, the one-dimensional case at $U/t=4$, $n=1$ and $L=10$ has been explored by
two-particle reduced density matrix (2-RDM)~\cite{hammond2006variational}
and many-body bootstrap~\cite{scheer2026bootstrapping},
and the best reported lower bound on the ground-state energy density still has an error of $5\times 10^{-3}$.
Using NGA bootstrap, we improve this bound gap by two orders of magnitude to $6\times 10^{-5}$ with moderate computational resources, as shown in Fig.~\ref{fig:hubbard}.
In Fig.~\ref{fig:hubbard}(a)(b) we compare the NGA bootstrap with manually selected basis hierarchies at $n=1$ and $L=10$.
The NGA algorithm expands the basis incrementally,
while manually selected sequences grow rapidly and soon become computationally inaccessible.
The NGA trajectory can depend sensitively on the choice of the initial basis.
For the NGA run, we start from $\mathcal{B}_0=\mathcal{S}^{0,0}_{5,2,5}$.
This choice is motivated by the observation that, during the NGA iteration,
the basis operators rapidly collapse into certain spin-resolved fermion-parity sector;
we therefore restrict the initial basis to a fixed fermion-parity sector from the outset.
It is found that the NGA basis is much more compact than the selected hierarchies
in the sense that fewer basis operators generate a larger SDP problem and yield stronger bootstrap bounds.
The final NGA basis contains 549 translation representatives, i.e. $\abs{\mathcal{B}_0}=549$,
compared with $|\mathcal{S}_{8,2,5}|=1036$ and $|\mathcal{S}_{6,3,2}|=2264$ for the final bases of the two manual sequences.
It also produces tighter energy lower bounds at comparable per-step solving time.
Moreover, as the NGA basis grows, the energy error decreases approximately as a power law in the SDP degrees of freedom and computation time,
without signs of saturation over our accessible range.
We further report the NGA bounds for different fillings and system sizes in Fig.~\ref{fig:hubbard}(c)(d).
%--------------------------------------------------------------------------

\section{Certified observables}
\begin{figure}[htbp]
    \centering\hspace{-1.25cm}
    \includegraphics[width=.9\columnwidth]{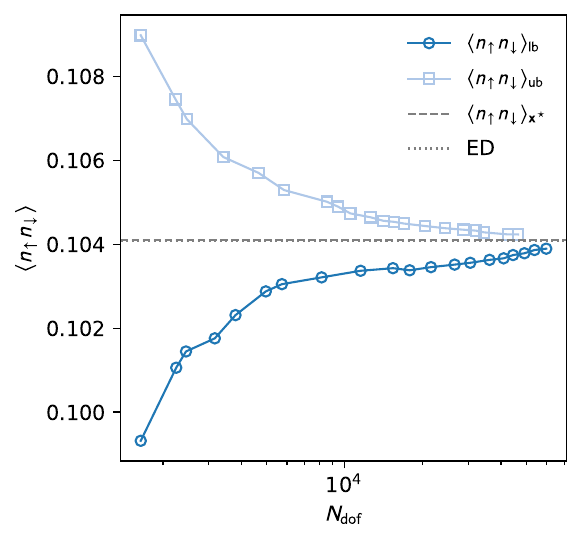}
    \caption{%
        Certified two-sided NGA bounds for the double occupancy of Hubbard chain at $U/t=4$, $n=1$ and $L=10$.
        The certified energy window is set according to the ED value and SDP lower bound in Fig.~\ref{fig:hubbard}(d).
        The exact double occupancy obtained from ED is $0.104085$
        and the uncertified SDP value $\expt[false]{n_\uparrow n_\downarrow}_{\bm{x}^\star}$ is $0.104105$.
        Our best certified bounds yield
        $\expt[false]{n_\uparrow n_\downarrow}_{\mathrm{lb}}=0.103900$ and
        $\expt[false]{n_\uparrow n_\downarrow}_{\mathrm{ub}}=0.104239$.
    }
    \label{fig:double_occ}
\end{figure}

%--------------------------------------------------------------------------
The NGA bootstrap also extends directly to give certified two-sided bounds on general observables.
For an observable $O$ with $\expt{O}=\bm{o}^T\bm{x}$,
we optimize $\bm{o}^T\bm{x}$ over the same relaxed feasible set supplemented by a certified energy window
$
    E_\mathrm{lb}\leq \bm{c}^T\bm{x}\leq E_\mathrm{ub}
$~\cite{han2020quantum,wang2024certifying}.
The resulting SDPs,
\begin{equation}
\begin{aligned}
    O&_\mathrm{lb/ub} = \mathop{\mathrm{min/max}}_{\bm{x}} \quad \bm{o}^T\bm{x} \\
    &\mathrm{s.t.} \quad M(\bm{x}) \succeq 0,\; A\bm{x}=\bm{b},\; E_\mathrm{lb} \leq \bm{c}^T\bm{x} \leq E_\mathrm{ub},
\end{aligned}
\end{equation}
yield certified bounds
$
    O_\mathrm{lb}\leq \expt{O}_0\leq O_\mathrm{ub}
$.
Applying the same NGA algorithm to these SDPs systematically tightens the two-sided bounds.
We briefly illustrate this process in Fig.~\ref{fig:double_occ} for the double occupancy of Hubbard chain.
The final observable bound gap depends on both the bootstrap basis and the imposed energy window.

%--------------------------------------------------------------------------

\section{Discussions}
%--------------------------------------------------------------------------
In this work, we introduced the NGA bootstrap, which improves the bootstrap bounds by iteratively inspecting the nullspace of the optimized moment matrix.
Although our numerical demonstrations focused on the ground state of finite-size one-dimensional systems,
the NGA framework can be naturally extended to systems
in the thermodynamic limit~\cite{han2020quantum},
in higher dimensions,
at finite temperatures~\cite{fawzi2024certified},
and with quenched disorder~\cite{onder2026bootstrapping}.
Moreover, the NGA bounds exhibit favorable scaling with the size of the bootstrap basis and the associated SDP,
while the SDP solving efficiency and memory consumption become the primary bottlenecks.
A promising direction is thus to develop dedicated solvers tailored to physical moment SDPs~\cite{baumgratz2012lower,simmonsduffin2015semidefinite,he2025qics,wang2025solving},
which may offer better efficiency and reduced memory cost compared with generic primal-dual interior-point methods.
In addition, the incremental nature of the NGA basis updates suggests that
part of the SDP construction and solving process may be reused between consecutive steps.
Exploiting this structure could further improve the overall efficiency of the NGA bootstrap.
In conclusion, we are optimistic that the NGA bootstrap will provide increasingly competitive certified bounds in larger systems and higher dimensions, and has the potential to complement state-of-the-art variational methods such as DMRG to provide accurate and reliable estimations to ground-state energy and observables for quantum many-body systems.
%--------------------------------------------------------------------------

\begin{acknowledgments}
%--------------------------------------------------------------------------
This work is supported by
the National Key R\&D Program of China (Grant No.~2022YFA1403402),
the National Natural Science Foundation of China (Grant No.~12174068),
the Science and Technology Commission of Shanghai Municipality (Grant Nos.~24LZ1400100 and 23JC1400600),
and the Shuguang Program of Shanghai Education Development Foundation and Shanghai Municipal Education Commission.
The code for the NGA bootstrap is available at \url{https://github.com/JefferyWangSH/QMBBoot-NGA}.
The numerical calculations were performed using computational resources provided by Hefei National Laboratory.
%--------------------------------------------------------------------------
\end{acknowledgments}

\appendix

\section{Symmetries in many-body bootstrap}\label{app:symmetries}
%--------------------------------------------------------------------------
\subsection{Symmetric density matrices}
Let $G$ be a finite or compact symmetry group of Hamiltonian $H$,
represented on the Hilbert space $\mathcal{H}$ by unitary or antiunitary operators $S_g$ satisfying $S_g^{-1}H S_g=H$ for all $g\in G$.
In the bootstrap formulation, we assume a $G$-symmetric density matrix and impose the associated symmetry constraints.
Starting from any ground-state density matrix $\rho_0$ in the target symmetry sector,
such a $G$-symmetric density matrix can be constructed via the group averaging
\begin{equation}
    \rho_{0,G} = \int_G \mathrm{d}g\, S_g^{-1}\rho_0 S_g,
\end{equation}
where $\mathrm{d}g$ is the normalized Haar measure~\cite{folland2015course} for compact continuous groups.
For a finite group, the integral is replaced by the normalized group sum.
Provided that the target symmetry sector is preserved by $G$,
each $S_g^{-1}\rho_0 S_g$ is a valid ground-state density matrix in the target symmetry sector,
and hence so is $\rho_{0,G}$.
By construction, it is symmetric under $G$ so that
\begin{equation}
    S_g^{-1} \rho_{0,G} S_g = \rho_{0,G},\quad \forall\, g\in G.
\end{equation}
Therefore, assuming $G$-symmetric density matrices in the many-body bootstrap does not exclude the target ground-state energy.
If the ground state is unique, the pure-state density matrix is already symmetric.
If the ground states are degenerate, a symmetric state $\rho_{0,G}$ can be constructed as above,
while possible symmetry-breaking order should be diagnosed through suitable $G$-invariant correlation functions.
For a $G$-symmetric density matrix, expectation values of any operator $O$ obey the symmetry constraints as discussed in the main text.

\subsection{Positive semidefinite (PSD) blocks}\label{sec:psd_block}
The memory and computational cost of the semidefinite programming (SDP) are largely governed by the dimension of the moment matrix $M$.
When the moment functional is $G$-symmetric and the finite operator space $\mathcal{V}$ is closed under unitary actions of $G$,
the moment matrix can be block diagonalized in its irreducible-representation basis,
so that the single PSD constraint on $M$ is equivalent to PSD constraints on smaller symmetry blocks.
In this subsection, we derive this block decomposition for general unitary symmetries.
The simplest antiunitary symmetry, complex conjugation, is discussed separately in Sec.~\ref{sec:cpx_conj}.

We assume that the finite operator space $\mathcal{V}=\mathrm{span}\,\mathcal{B}$ is closed under unitary symmetry actions,
i.e. $U_g^{-1} O U_g\in\mathcal{V}$ for all $O\in\mathcal{V}$ and $g\in G$,
so that the conjugation by $U_g$ defines a unitary representation on $\mathcal{V}$.
Then $\mathcal{V}$ can be decomposed into irreducible representation (irrep) subspaces $\mathcal{V}_\lambda$ of $G$ as
\begin{equation}
    \mathcal{V}=\bigoplus_\lambda \bigoplus_{a=1}^{m_\lambda} \mathcal{V}_{\lambda,a}.
\end{equation}
Here $\lambda$ labels the irreducible representation,
$a=1,\ldots,m_\lambda$ labels copies of the same representation,
and $m_\lambda$ is the multiplicity.
Accordingly, the operators $O_i\in\mathcal{V}$ can be linearly recombined into multiplets $O_{\lambda,a,\alpha}$ that transform irreducibly under $G$,
\begin{equation}
    U_g^{-1}O_{\lambda,a,\alpha} U_g
    = \sum_\beta D^{(\lambda)}_{\beta\alpha}(g) O_{\lambda,a,\beta},
\end{equation}
where $D^{(\lambda)}(g)$ is the representation matrix of $g$ in the irreducible representation $\lambda$,
and $\alpha,\beta=1,\ldots,\dim\mathcal{V}_\lambda$ label the components within the irreducible representation.

Because $\rho$ is $G$-symmetric, the moments are invariant under the action of $U_g$,
\begin{equation}\begin{aligned}
    M_{\lambda a\alpha,\mu b\beta}
    &=
    \left\langle
    O_{\lambda,a,\alpha}^\dag O_{\mu,b,\beta}
    \right\rangle \\[5pt]
    &=
    \sum_{\alpha'\beta'}
    D^{(\lambda)}_{\alpha'\alpha}(g)^\ast
    M_{\lambda a\alpha',\mu b\beta'}
    D^{(\mu)}_{\beta'\beta}(g),
\end{aligned}\end{equation}
or compactly,
\begin{equation}
    M_{\lambda a,\mu b} = D^{(\lambda)}(g)^\dag M_{\lambda a,\mu b} D^{(\mu)}(g), \quad \forall\, g\in G.
\end{equation}
The block $M_{\lambda a,\mu b}$ defines a $G$-equivariant linear map from $\mathcal{V}_{\mu,b}$ to $\mathcal{V}_{\lambda,a}$.
By Schur's lemma, blocks connecting inequivalent irreducible representations $\lambda$ and $\mu$ must vanish,
whereas blocks with $\lambda=\mu$ are proportional to the identity on the irrep indices, i.e.
\begin{equation}
    M_{\lambda a\alpha,\mu b\beta}
    = \left[M_{\lambda a,\mu b}\right]_{\alpha,\beta}
    = \delta_{\lambda\mu} \left[A_\lambda\right]_{ab} \delta_{\alpha\beta},
\end{equation}
or equivalently,
\begin{equation}
    M = \bigoplus_\lambda \left(A_\lambda\otimes I_{\dim \mathcal{V}_\lambda}\right),
\end{equation}
where $A_\lambda$ acts on the multiplicity space.
Thus the original PSD constraint on $M$ is equivalent to PSD constraints on the smaller blocks $A_\lambda$.
This irrep-level block diagonalization requires the finite operator space $\mathcal{V}$ to be closed under the symmetry actions,
such that $G$ defines a representation on $\mathcal{V}$.
In an NGA bootstrap, however, this closure is not automatically guaranteed
because NGA moves do not necessarily keep complete symmetry orbits.

For Abelian symmetries, the block structure is especially simple
because every irreducible representation is one-dimensional and specified by a character,
\begin{equation}\label{eq:abelian_rep}
    U_g^{-1} O_{\lambda,a} U_g = \chi_\lambda(g) O_{\lambda,a}.
\end{equation}
Then
\begin{equation}\label{eq:abelian_block}
    M = \bigoplus_\lambda A_\lambda.
\end{equation}
Here $M_{\lambda a,\mu b}=\expt[false]{O^\dag_{\lambda,a} O_{\mu,b}}$ denotes the moment matrix after reorganizing the basis operators by Abelian charges,
and $A_\lambda$ is an $m_\lambda\times m_\lambda$ Hermitian matrix.
In this work, we use lattice translation group $\mathbb{Z}_L$, Sec.~\ref{sec:translation}, to decompose the PSD constraint into momentum PSD blocks for both Ising and Hubbard chain.
For Hubbard chain, spin-resolved fermion parity and particle-hole parity, specific at half-filling, are used to further decompose each momentum PSD block into parity blocks, as discussed in Sec.~\ref{sec:fermion_parity} and Sec.~\ref{sec:ph_eta}.

\subsection{Lattice translation}\label{sec:translation}
For a periodic chain of length $L$, the lattice translations form an Abelian group $T = \{T(s),\ s=0,\dots,L-1\}$,
whose irreducible representations are one-dimensional and labeled by momentum $k=2\pi n/L$, $n=0,\ldots,L-1$ with characters $\chi_k(s)=e^{iks}$.
In practice, the NGA bootstrap basis $\mathcal{B}$ involves all translation orbits.
We first choose a set of translation representatives,
\begin{equation}\label{eq:bootstrap_basis_reprs}
    \mathcal{B}_0 = \left\{O_a(0):a=1,\ldots,N_a\right\}.
\end{equation}
The full bootstrap basis $\mathcal{B}$ is then constructed by translating $\mathcal{B}_0$,
\begin{equation}\begin{aligned}
    \label{eq:bootstrap_basis}
    \mathcal{B} &= \bigcup_{r=0}^{L-1} \mathcal{B}(r),\\[5pt]
    \mathcal{B}(r) &= \left\{T^\dag(r)O_a(0)T(r): a=1,\ldots,N_a\right\}.
\end{aligned}\end{equation}
The irrep basis is obtained by Fourier transformation,
\begin{equation}
    O_a(k) = \frac{1}{\sqrt L} \sum_{r=0}^{L-1} e^{-ikr} O_a(r),
\end{equation}
where we define $O_a(r)=T^\dag(r)O_a(0)T(r)$.
With these conventions,
\begin{equation}
    T^\dag(s) O_a(k) T(s) = e^{iks} O_a(k),
\end{equation}
so $O_a(k)$ carries Abelian charge $k$.
We define the moment matrix by $M_{ar,bs}=\expt[false]{O_a^\dagger(r)O_b(s)}$.
According to Eqs.~\eqref{eq:abelian_rep} and \eqref{eq:abelian_block},
the moment matrix decomposes as
\begin{equation}
    F^\dag M F = \bigoplus_k M(k).
\end{equation}
Therefore, the PSD constraint $M\succeq 0$ is replaced equivalently by $M(k)\succeq 0$ for all $k$.
The block entries are
\begin{equation}
    M_{ab}(k) = \sum_{r=0}^{L-1}e^{ikr}\left\langle O_a^\dag(r)O_b(0)\right\rangle.
\end{equation}
We note that larger system size $L$ provides linearly more momentum PSD blocks, whose dimension is determined by $N_a$.
Additional symmetries that commute with translations and preserve the operator space can further decompose each $M(k)$ block.

\subsection{Complex conjugation}\label{sec:cpx_conj}
Complex conjugation $\mathcal{K}$ provides the simplest example of an antiunitary symmetry.
It is defined only after choosing a computational basis $\ket{n}$ of the Hilbert space,
\begin{equation}
    \mathcal{K}\ket{\psi} = \sum_n \psi_n^\ast\ket{n}.
\end{equation}
Thus $\mathcal{K}$ is antilinear and satisfies $\mathcal{K}^{-1}i\mathcal{K} = -i$ and $\mathcal{K}^2=1$.
For spin systems, we use the product basis of local $Z$ eigenstates as the computational basis so that
\begin{equation}\label{eq:pauli_k_parity}
    \mathcal{K}^{-1} X_j \mathcal{K} = X_j, \quad
    \mathcal{K}^{-1} Y_j \mathcal{K} = -Y_j, \quad
    \mathcal{K}^{-1} Z_j \mathcal{K} = Z_j.
\end{equation}
For fermions, we use the occupation-number basis in which $c$ and $c^\dag$ have real matrix elements, and hence
\begin{equation}
    \mathcal{K}^{-1} c_j \mathcal{K} = c_j, \quad
    \mathcal{K}^{-1} c^\dag_j \mathcal{K} = c^\dag_j.
\end{equation}
Equivalently for Majoranas $\gamma^1_j=c^\dag_j+c_j$ and $\gamma^2_j=i(c^\dag_j-c_j)$, we have
\begin{equation}\label{eq:majorana_k_parity}
    \mathcal{K}^{-1} \gamma^1_j \mathcal{K} = \gamma^1_j, \quad
    \mathcal{K}^{-1} \gamma^2_j \mathcal{K} = -\gamma^2_j.
\end{equation}

Assume that complex conjugation is a symmetry of the Hamiltonian
and the bootstrap basis $\mathcal{B}$ in Eq.~\eqref{eq:bootstrap_basis} is closed under $\mathcal{K}$.
Moreover, $\mathcal{K}$ commutes with lattice translation, and we write
\begin{equation}
    \mathcal{K}^{-1} O_a(r) \mathcal{K} = \sum_b C_{ba} O_b(r).
\end{equation}
Because the induced map $O\mapsto \mathcal{K}^{-1} O \mathcal{K}$ is antilinear instead of linear,
$C$ is not an ordinary linear representation
and should be understood merely as the coefficient matrix describing these actions on the basis operators.
In addition, $C$ is by construction invertible with $C^{-1}=C^\ast$ since $\mathcal{K}^2=1$.
Since $\mathcal{K}$ complex conjugates the Fourier phase, the momentum operators satisfy
\begin{equation}
    \mathcal{K}^{-1} O_a(k) \mathcal{K} = \sum_b C_{ba} O_b(-k).
\end{equation}
Therefore, complex conjugation maps the momentum sector $k$ to $-k$.
For a $\mathcal{K}$-symmetric density matrix, antiunitary symmetry gives $\expt[false]{O}=\expt[false]{\mathcal{K}^{-1} O \mathcal{K}}^\ast$.
Applying this to $O^\dag_a(k) O_b(k)$ yields
\begin{equation}\begin{aligned}
    M_{ab}(k)
    &= \expt{O^\dag_a(k) O_b(k)}
    = \expt{\mathcal{K}^{-1} O^\dag_a(k) O_b(k) \mathcal{K}}^\ast\\[5pt]
    &= \sum_{a'b'} C_{a'a} C^\ast_{b'b} M^\ast_{a'b'}(-k).
\end{aligned}\end{equation}
Equivalently,
\begin{equation}\label{eq:momentum_block_equiv}
    M(k) = \left(C^\ast\right)^\dag M^\ast(-k)\, C^\ast.
\end{equation}
This establishes an invertible congruence transformation connecting $M^\ast(-k)$ and $M(k)$,
which implies
\begin{equation}
    M(k)\succeq 0 \iff M^\ast(-k)\succeq 0 \iff M(-k)\succeq 0.
\end{equation}
The last equivalence is inferred from the fact that $M(k)$ is Hermitian.
Therefore the momentum PSD constraints in the opposite momentum sectors are equivalent.
For momenta $k^\star$ satisfying $k^\star=-k^\star$ modulo $2\pi$,
namely $k^\star=0$ for any $L$ and $k^\star=\pi$ only when $L$ is even,
Eq.~\eqref{eq:momentum_block_equiv} becomes an additional reality constraint within the same momentum block.

In this work, we choose the translation representatives $O_a$ to have definite $\mathcal{K}$ parity.
For spin models, $O_a$ are individual Pauli strings $P_a$, i.e. products of single-site Pauli operators as in Eq.~\eqref{eq:pauli_string}.
For fermionic models, $O_a$ are canonical Majorana monomials $\Gamma_a$, i.e. products of local Majorana modes as in Eq.~\eqref{eq:majorana_monomial}.
As inferred from Eqs.~\eqref{eq:pauli_k_parity} and \eqref{eq:majorana_k_parity},
these operators are eigenoperators of $\mathcal{K}$ in their corresponding computational bases, e.g.
\begin{equation}
    \mathcal{K}^{-1} P_a \mathcal{K} = \kappa_a P_a,
    \quad
    \kappa_a = (-1)^{N_y}
\end{equation}
for spin models with $N_y$ the number of $Y$ operators in $P_a$ and
\begin{equation}
    \mathcal{K}^{-1} \Gamma_a \mathcal{K} = \kappa_a \Gamma_a,
    \quad
    \kappa_a = (-1)^{N_2}
\end{equation}
for fermionic models.
$N_2$ counts the number of $\gamma^2$ operators in $\Gamma_a$.
With these conventions, the matrix $C$ becomes diagonal, $C_{ba}=\kappa_a\delta_{ba}$ with $\kappa_a=\pm1$,
and we have
\begin{equation}
    M_{ab}(k) = \kappa_a \kappa_b M_{ab}(-k)^\ast.
\end{equation}
For momenta $k^\star$ that are invariant under $k\mapsto -k$, this reveals the following reality structure of $M(k^\star)$,
\begin{equation}
    M(k^\star) =
    \begin{pmatrix}
    M_1 & iM_3\\
    -iM_3^T & M_2\\
    \end{pmatrix}
    .
\end{equation}
Here $M_1$, $M_2$ are real symmetric matrices and $M_3$ is a general real matrix.
We have reordered $\mathcal{B}(k^\star)=\{O_a(k^\star)\}$ as $\mathcal{B}_+(k^\star)\oplus\mathcal{B}_-(k^\star)$ according to their $\mathcal{K}$ parity.
Therefore the original complex PSD constraint $M(k^\star)\succeq 0$ is equivalent to a real PSD constraint given by
\begin{equation}
    \tilde{M}(k^\star) = U^\dag M(k^\star) U
    =
    \begin{pmatrix}
    M_1 & -M_3\\
    -M_3^T & M_2\\
    \end{pmatrix}
    \succeq 0,
\end{equation}
with
\begin{equation}
    U =
    \begin{pmatrix}
    I & 0\\
    0 & iI\\
    \end{pmatrix}
    .
\end{equation}
%--------------------------------------------------------------------------

\section{Ising chain}
%--------------------------------------------------------------------------
We bootstrap the Ising chain with Hamiltonian
\begin{equation}
    H = -J \sum_i Z_i Z_{i+1} - h \sum_i X_i - h_z \sum_i Z_i.
\end{equation}
The translation-representative operator basis $\mathcal{B}_0$ in Eq.~\eqref{eq:bootstrap_basis_reprs} consists of individual Pauli strings of the form
\begin{equation}\label{eq:pauli_string}
    P_a = \sigma^{\alpha_1}_{x_1} \sigma^{\alpha_2}_{x_2} \cdots \sigma^{\alpha_d}_{x_d}.
\end{equation}
Each $\sigma^{\alpha_j}_{x_j}$ denotes a Pauli $X$, $Y$, or $Z$ operator at site $x_j$ and $d$ is the degree of Pauli string, i.e. the number of nontrivial local Pauli operators.
We use translation symmetry to block diagonalize the moment matrix and complex conjugation to identify the equivalent $k$ and $-k$ momentum sectors,
as the full bootstrap basis $\mathcal{B}$ is by construction closed under these symmetry actions.
Furthermore, lattice inversion $\mathcal{I}: x \mapsto -x$ is utilized to reduce SDP variables according to
\begin{equation}
    \expt{O} = \expt{\mathcal{I}^{-1} O \mathcal{I}}.
\end{equation}
We also impose the stationarity constraints $\expt[false]{[H,O]}=0$ whenever it is representable by the SDP variable $\bm{x}$.
%--------------------------------------------------------------------------

\section{Hubbard chain}\label{app:hubbard}
%--------------------------------------------------------------------------
The Hamiltonian of Hubbard chain in the particle-hole symmetric form is
\begin{equation}\begin{aligned}
    \label{eq:hubbard_hamil}
    H = &-t \sum_{i\sigma} \left(c^\dag_{i,\sigma} c_{i+1,\sigma} + c^\dag_{i+1,\sigma} c_{i,\sigma}\right)\\
    &+ U \sum_i \left(n_{i\uparrow}-\frac{1}{2}\right) \left(n_{i\downarrow}-\frac{1}{2}\right).
\end{aligned}\end{equation}
We introduce Majorana fermion operators, $\gamma^1_{i\sigma}=c^\dag_{i\sigma}+c_{i\sigma}$ and $\gamma^2_{i\sigma}=i(c^\dag_{i\sigma}-c_{i\sigma})$,
where $\gamma^\alpha_{i\sigma}$ are Hermitian and obey the Clifford algebra $\{\gamma^\alpha_{i\sigma},\gamma^\beta_{j\sigma'}\}=2\delta_{\alpha\beta}\delta_{ij}\delta_{\sigma\sigma'}I$.
In terms of Majorana fermions, the Hamiltonian becomes
\begin{equation}\begin{aligned}
    H &= \frac{t}{2} \sum_{i\sigma} \left(-i \gamma^1_{i,\sigma} \gamma^2_{i+1,\sigma} + i \gamma^2_{i,\sigma} \gamma^1_{i+1,\sigma}\right)\\
    &- \frac{U}{4} \sum_{i} \gamma^1_{i\uparrow} \gamma^2_{i\uparrow} \gamma^1_{i\downarrow} \gamma^2_{i\downarrow}.
\end{aligned}\end{equation}
The local Hilbert space of Hubbard model is four-dimensional such that the local operator algebra has dimension 16,
spanned by the $2^4$ canonical Majorana monomials generated from the four local Majorana modes $\gamma^\alpha_{i\sigma}$ at site $i$.
All operators in the translation-representative basis $\mathcal{B}_0$ are canonical Majorana monomials of the form
\begin{equation}\label{eq:majorana_monomial}
    \Gamma_a = \gamma^{\alpha_1}_{x_1,\sigma_1} \gamma^{\alpha_2}_{x_2,\sigma_2} \cdots \gamma^{\alpha_d}_{x_d,\sigma_d}.
\end{equation}
In each $\Gamma_a$, the Majorana operators are distinct and arranged in the canonical order,
defined by sorting $(x_j,\sigma_j,\alpha_j)$ in ascending order from left to right.
$d$ denotes the degree of the monomial.
Compared with a complex-fermion basis, the Majorana basis is more convenient because
the product of two Majorana monomials can be reduced to the canonical form simply by bit-wise operations plus a sign factor $\pm 1$ from reordering,
whereas normal-ordering complex-fermion products generally produce additional contraction terms.
This makes products and commutators of Majorana monomials particularly efficient to evaluate,
which is important because these algebraic operations dominate the computational cost of SDP compilation and NGA basis growth.

We restrict the bootstrap to a fixed particle-number sector.
Let $N = \sum_{i,\sigma} n_{i\sigma}$ and $N_0$ be the target particle number.
We impose linear constraints
\begin{equation}
    \expt{N-N_0I}=0, \quad \expt{(N-N_0I)^2}=0.
\end{equation}
The first condition fixes the mean particle number,
while the second forces its variance to vanish,
thereby restricting the state to the particle-number sector with fixed $N_0$.

In addition to the translation, lattice inversion, and complex conjugation symmetry used in the Ising chain,
we use more dedicated symmetry reductions and constraints for the Hubbard model.
The appropriate implementation of a symmetry depends on
how the symmetry acts on individual Majorana monomials in the truncated bootstrap basis,
which generally falls into three categories:
\begin{itemize}
\item \textit{PSD block decompositions.}
When the monomials have a definite symmetry charge, e.g. fermion parity or particle-hole parity at half filling,
the bootstrap basis is automatically closed under symmetry actions and the PSD constraints can be block diagonal as in Sec.~\ref{sec:psd_block}.
If the symmetry commutes with one-site translation,
each momentum PSD block can be further decomposed into associated charge blocks.
Otherwise, the symmetry maps one momentum sector to another and identifies equivalent momentum blocks.

\item \textit{Ward identities.}
Continuous unitary symmetries,
such as time translation, charge $U(1)$, spin $SU(2)$, and $\eta$-pairing $SU(2)_\eta$ at half-filling,
are imposed through Ward identities.
For their conserved charge $Q$, we impose
\begin{equation}
    \expt{[Q,O]} = 0
\end{equation}
whenever the commutator can be represented within the truncated moment space $\mathcal{M}$.

\item \textit{Direct reductions of SDP variables.}
For discrete symmetries whose action is not diagonal on the monomial basis,
or for which the current NGA basis is not closed under the full symmetry action,
we use symmetry relations of the form
\begin{equation}
    \expt{O} = \expt{U^{-1} O U}
\end{equation}
to prune equivalent SDP variables.
We apply this strategy to lattice inversion, spin exchange, and $C_4$ rotations in the Majorana plane.
From the perspective of the SDP solver,
eliminating redundant SDP variables is often more efficient than enforcing the same relations through explicit linear constraints.
\end{itemize}
Below, we describe these symmetries in the Hubbard chain and the corresponding symmetry reductions and constraints used in the bootstrap.

\subsection{Spin-resolved fermion parity}\label{sec:fermion_parity}
Consider the fermion-parity symmetry of the two spin species,
\begin{equation}
    \mathcal{P}_\uparrow=(-1)^{N_\uparrow}, \quad
    \mathcal{P}_\downarrow=(-1)^{N_\downarrow},
\end{equation}
which generates a $\mathbb{Z}_{2,\uparrow}\times\mathbb{Z}_{2,\downarrow}$ subgroup of $U_\uparrow(1)\times U_\downarrow(1)$.
Each Majorana monomial has a definite fermion-parity charge $(p_\uparrow,p_\downarrow)$,
determined by the numbers of up- and down-spin Majoranas modulo two.
Since $\mathcal{P}_\uparrow$ and $\mathcal{P}_\downarrow$ commute with the lattice translation,
the operator space $\mathcal{V}$ can be simultaneously decomposed into sectors labeled by momentum and fermion parity.
We therefore organize each momentum basis as
\begin{equation}
    \mathcal{B}(k) = \bigoplus_{p_\uparrow,p_\downarrow=0,1} \mathcal{B}_{p_\uparrow,p_\downarrow}(k),
\end{equation}
where $\mathcal{B}_{p_\uparrow,p_\downarrow}(k)$ consists of monomials with fermion parity $(p_\uparrow,p_\downarrow)$.
According to Eqs.~\eqref{eq:abelian_rep} and \eqref{eq:abelian_block},
each momentum PSD block decomposes as
\begin{equation}
    M(k) = \bigoplus_{p_\uparrow,p_\downarrow} M_{p_\uparrow,p_\downarrow}(k).
\end{equation}

\subsection{Charge and spin rotations}
The Hubbard model has continuous charge $U(1)$ and spin $SU(2)$ symmetries.
The corresponding generators are the total particle number $N=N_\uparrow+N_\downarrow$
and the total spin operators $S^+$, $S^-$, and $S^z$.
At the Lie-algebra level, the pair $(N_\uparrow,N_\downarrow)$ spans the same subalgebra as $(N,S^z)$.
We therefore implement the charge and $S^z$ Ward identities using $N_\uparrow$ and $N_\downarrow$ through
\begin{equation}
    \expt{[N_\uparrow,O]} = 0, \quad \expt{[N_\downarrow,O]} = 0.
\end{equation}
The Ward identities for transverse spin rotations can equivalently be imposed using the ladder operators $S^\pm$.
Since $(S^+)^\dag=S^-$, and a canonical Majorana monomial satisfies $O^\dag=\chi_O O$ with $\chi_O=\pm1$, we have
\begin{equation}
    \left[S^-,O\right] = -\chi_O \left[S^+,O\right]^\dag.
\end{equation}
This implies that the Ward identities from $S^+$ and $S^-$ are equivalent under complex conjugation.
We therefore keep only the $S^+$ Ward identities,
\begin{equation}
    \expt{[S^+,O]} = 0,
\end{equation}
whenever the commutator is representable by the current SDP variables.

\subsection{Spin exchange}
The spin exchange operation combines a spin $\pi$-rotation around $x$ with a charge $U(1)$ rotation,
\begin{equation}
    \mathcal{X}_s = e^{-i\pi N/2} e^{i\pi S^x}.
\end{equation}
It exchanges the spin labels of Majorana operators as
\begin{equation}
    \mathcal{X}_s^{-1} \gamma^\alpha_{i\uparrow} \mathcal{X}_s = \gamma^\alpha_{i\downarrow},
    \quad
    \mathcal{X}_s^{-1} \gamma^\alpha_{i\downarrow} \mathcal{X}_s = \gamma^\alpha_{i\uparrow}.
\end{equation}
Unlike fermion parity, the action of spin exchange is not diagonal on individual Majorana monomials,
and the NGA basis is not guaranteed to be closed under spin exchange.
Therefore we use 
\begin{equation}
    \expt{O} = \expt{\mathcal{X}_s^{-1} O \mathcal{X}_s}
\end{equation}
to prune equivalent SDP variables.

\subsection{\texorpdfstring{$C_4$}{C4} rotation in the Majorana plane}
For each spin species, the Majorana spinor
\begin{equation}
    \boldsymbol{\gamma}_{i\sigma} = 
    \begin{pmatrix}
        \gamma^1_{i\sigma}\\[5pt]
        \gamma^2_{i\sigma}
    \end{pmatrix}
\end{equation}
forms a real two-dimensional Majorana plane. The $U_\sigma(1)$ charge rotation generated by $N_\sigma$ with spin $\sigma$ acts as an $SO(2)$ rotation in this plane that
\begin{equation}
    U_\sigma(\theta)^{-1} \boldsymbol{\gamma}_{i\sigma} U_\sigma(\theta)
    = R(\theta) \boldsymbol{\gamma}_{i\sigma},
    \quad
    R(\theta)
    =
    \begin{pmatrix}
        \cos\theta & \sin\theta\\
        -\sin\theta & \cos\theta
    \end{pmatrix},
\end{equation}
where $U_\sigma(\theta) = e^{-i\theta N_\sigma}$.
We remark that the continuous $U_\sigma(1)$ symmetry has been imposed through the Ward identities generated by $N_\sigma$, implemented as linear constraints among SDP variables.
Here, for the direct reduction of SDP variables, we use the finite $C_{4,\sigma}$ subgroup generated by the quarter rotation $\theta=\pi/2$ so that
\begin{equation}
    \expt{O} = \expt{\mathcal{R}^{-1} O \mathcal{R}}, \quad
    \mathcal{R} \in C_{4,\uparrow} \times C_{4,\downarrow}.
\end{equation}

\subsection{Particle-hole symmetry and \texorpdfstring{$\eta$}{eta}-pairing at half-filling}\label{sec:ph_eta}
The Hubbard Hamiltonian in Eq.~\eqref{eq:hubbard_hamil} is written in the particle-hole symmetric form.
For an even-length periodic chain, the lattice is bipartite
and $H$ is invariant under the particle-hole (PH) symmetry defined by
\begin{equation}
    \mathcal{P}_\mathrm{ph}^{-1} c_{i\sigma} \mathcal{P}_\mathrm{ph} = \epsilon_i c^\dag_{i\sigma}, \quad
    \mathcal{P}_\mathrm{ph}^{-1} c^\dag_{i\sigma} \mathcal{P}_\mathrm{ph} = \epsilon_i c_{i\sigma},
\end{equation}
with the staggered sign $\epsilon_i=(-1)^i$.
We remark that a PH-symmetric density matrix is incompatible with fixed-filling constraints away from half filling,
since the PH transformation maps the total particle number as $N\mapsto 2L-N$.
Therefore, we impose PH symmetry only at half filling.

The PH transformation acts on Majorana operators as
\begin{equation}
    \mathcal{P}_\mathrm{ph}^{-1} \gamma^1_{i\sigma} \mathcal{P}_\mathrm{ph} = \epsilon_i \gamma^1_{i\sigma}, \quad
    \mathcal{P}_\mathrm{ph}^{-1} \gamma^2_{i\sigma} \mathcal{P}_\mathrm{ph} = -\epsilon_i \gamma^2_{i\sigma}.
\end{equation}
Hence each real-space Majorana monomial has definite PH parity.
However, $\mathcal{P}_\mathrm{ph}$ does not generally commute with one-site translation because of the staggered factor.
If monomial $O_a$ has PH parity $p_{\mathrm{ph},a}$, then
\begin{equation}
    \mathcal{P}_\mathrm{ph}^{-1} O_a(r) \mathcal{P}_\mathrm{ph} = (-1)^{\,\abs{O_a}r} (-1)^{p_{\mathrm{ph},a}} O_a(r),
\end{equation}
where $\abs{O_a}$ is the degree of the monomial.
For the momentum operator, this gives
\begin{equation}\label{eq:ph_transform}
    \mathcal{P}_\mathrm{ph}^{-1} O_a(k) \mathcal{P}_\mathrm{ph} = (-1)^{p_{\mathrm{ph},a}} O_a\left(k+\pi\abs{O_a}\right).
\end{equation}
Therefore the effects of PH should be considered separately for basis operators $O_a$ with even and odd degree.
Suppose that the PSD matrix has been decomposed into momentum and fermion-parity blocks with operator basis $\mathcal{B}_{p_\uparrow,p_\downarrow}(k)$.
The operators in $\mathcal{B}_{p_\uparrow,p_\downarrow}(k)$ have even degree if $p_\uparrow=p_\downarrow$ and have odd degree otherwise.
For PSD blocks with $p_\uparrow=p_\downarrow$,
Eq.~\eqref{eq:ph_transform} acts internally for each momentum $k$,
and $M_{p_\uparrow, p_\downarrow}(k)$ can be further decomposed into PH-parity blocks as
\begin{equation}
    M_{p_\uparrow, p_\downarrow}(k) = \bigoplus_{p_\mathrm{ph}=0,1} M_{p_\uparrow, p_\downarrow, p_\mathrm{ph}}(k), \quad p_\uparrow=p_\downarrow.
\end{equation}
Instead, for PSD blocks with $p_\uparrow\neq p_\downarrow$,
the PH transformation in Eq.~\eqref{eq:ph_transform} maps $k$ to $k+\pi$ such that
\begin{equation}
    M_{p_\uparrow, p_\downarrow}(k) = D^\dag\, M_{p_\uparrow, p_\downarrow}(k+\pi)\, D, \quad p_\uparrow\neq p_\downarrow,
\end{equation}
where $D$ is the diagonal sign matrix with $D_{aa}=(-1)^{p_{\mathrm{ph},a}}$.
This establishes an equivalence between PSD constraints at momenta shifted by $\pi$ in sectors with $p_\uparrow\neq p_\downarrow$.
In summary, PH symmetry decomposes the PSD blocks with $p_\uparrow=p_\downarrow$ further into PH-parity blocks,
while, for $p_\uparrow\neq p_\downarrow$, it makes the PSD constraints at momenta separated by $\pi$ equivalent.

Moreover, on the bipartite lattice, the PH-symmetric Hubbard Hamiltonian also has the $\eta$-pairing, or pseudospin, $SU(2)_\eta$ symmetry~\cite{yang1989eta,yang1990so4} generated by
\begin{equation}
    \eta^+ = \sum_i \epsilon_i c^\dag_{i\uparrow}c^\dag_{i\downarrow}, \quad
    \eta^- = \left(\eta^+\right)^\dag, \quad
    \eta^z = \frac{1}{2} \left(N-LI\right).
\end{equation}
Again, an $SU(2)_\eta$-symmetric state cannot select a nonzero $\eta^z$ direction,
so that the $\eta$-pairing symmetry can be imposed through Ward identities only at half-filling.
In particular, we impose
\begin{equation}
    \expt{[\eta^+,O]} = 0.
\end{equation}
The constraints generated by $\eta^-$ are redundant with those from $\eta^+$ as in the spin $SU(2)$ case,
while the $\eta^z$ Ward identities are already covered by those induced by the charge $U(1)$.
%--------------------------------------------------------------------------

\bibliography{ref.bib}

%--------------------------------------------------------------------------
\end{document}